\documentclass[aps,prl,twocolumn,superscriptaddress,reprint]{revtex4-1}

\usepackage{float}
\usepackage{graphicx}
\usepackage{amsmath}
\usepackage{bm}
\usepackage{color}
\usepackage{amssymb}
\usepackage{mathrsfs}
\usepackage{dcolumn}
\usepackage{multirow}
\usepackage[mathlines]{lineno}
\usepackage[colorlinks=true,linkcolor=blue,urlcolor=blue,citecolor=blue]{hyperref}

\begin{document}
	
	
	\title{Interaction-Driven Topological Switch in a $P$-Band Honeycomb Lattice}
	
	\author{Hua Chen}
	\email{Electronic address: hwachanphy@zjnu.edu.cn} 
	\affiliation{Department of Physics, Zhejiang Normal University, Jinhua 321004, China}
	
	\author{X. C. Xie}
	\affiliation{International Center for Quantum Materials, School of Physics, Peking University, Beijing 100871, China}
	\affiliation{Collaborative Innovation Center of Quantum Matter, Beijing 100871, China}
	\affiliation{CAS Center for Excellence in Topological Quantum Computation, University of Chinese Academy of Sciences, Beijing 100190, China}		
		
	\begin{abstract}
	The non-interacting band structure of spinless fermions in a two-dimensional ($d=2$) $p$-band honeycomb lattice exhibits two quadratic band touching points (QBTPs), which lie at the Fermi levels of filling $\nu=1/4$ and its particle-hole conjugated filling $\nu=3/4$. A weak Hubbard interaction $U$ spontaneously breaks the time-reversal symmetry and removes the QBTP, rendering the system into a quantum anomalous Hall insulator (QAHI). The first-order topological nature of QAHI is characterized by a nontrivial Chern number and supports ($d-1$)-dimensional chiral edge modes. With increasing the interaction $U$, the system is driven into a Dirac semimetal by breaking the crystal symmetry through a discontinuous quantum phase transition. The emergent Dirac points each with Berry flux $\pi$ are generated in pairs, originating from the $2\pi$ Berry flux of QBTP. A sufficiently large $U$ ultimately drives the system into a dimerized insulator (DI) by simultaneously annihilating the Dirac points at the Brillouin zone boundary. The second-order topological nature of DI is characterized by the quantized polarizations and supports ($d-2$)-dimensional corner states. Our study provides a unique setting for exploring the topological switch between the first-order and second-order topological insulators.  
	\end{abstract}
	
	\date{\today}
	
	\maketitle
	

	Topological insulators (TIs) have recently received renewal research interests since the discovery of higher-order TIs~\cite{Benalcazar17-1,Benalcazar17-2}.
	An $n$th-order topological insulator in $d$ spatial dimensions is predicted to have topologically protected gapless or in-gap states that localized at the ($d-n$)-dimensional boundaries according to the bulk-boundary correspondence. As a paradigmatic example in two dimensions ($d=2$), the conventional TI exhibits one-dimensional gapless edge states and is thus categorized into the first-order TI~\cite{Hasan10,Qi11,Bernevig13}. While, the second-order TI instead exhibits zero-dimensional in-gap states at its corners. Symmetry, on the other hand, manifests its fundamental role in the Altland-Zirnbauer classification on first-order TIs~\cite{Altland97,Schnyder08,Kitaev09,Ryu10,Chiu16} as well as its extension on higher-order TIs~\cite{Langbehn17,Song17,Schindler18,Geier18}. Of particular interest is that the symmetry of correlated system can be altered by many-body interactions through spontaneous symmetry breaking and therefore provides a promising mechanism  for changing its topology. An interacting system with distinct broken symmetries can be classified into different topological classes or topological categories and may support completely different gapless or in-gap states at its boundaries.

	Here we study the interacting spinless fermions in a $p$-band honeycomb lattice and report an interaction-driven topological switch from a quantum anomalous Hall insulator (QAHI) 
	to a dimerized second-order topological insulator (DSOTI) with an intermediate Dirac semimetal (DSM).
	The QAHI with broken time-reversal symmetry shows a zero-field quantized Hall conductance $e^2/h$ and supports one-dimensional gapless chiral edge states~\cite{Klitzing80,Thouless82,Kohmoto85}. The QAHI is therefore identified as a first-order TI. In contrast, the DSOTI spontaneously breaks the crystal symmetry~\cite{Wu07}. Its second-order topology is further revealed by the quantized polarizations through the Wannier-Bloch duality between real and momentum space, supporting zero-dimensional corner states. Our finding, the topological switch between first-order and second-order TIs, broadens the concept of Landau's theory of symmetry breaking in describing phase transitions~\cite{Landau37-1,Landau37-2,Landau} and enriches the physics behind the interplay of symmetry and topology.

	\begin{figure}
		\centering
		\includegraphics[width=0.48\textwidth]{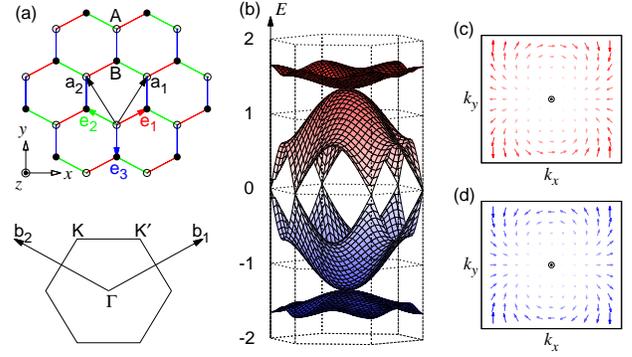}
		\caption{(color online)
			(a) The bipartite structure of honeycomb lattice and the hexagonal Brillouin zone. 
			(b) The band structure of tight-binding model in Eq.~(\ref{eq:TB}) with $\{t_\sigma,t_\pi\}=\{1,-0.1\}$.
			The pseudovector field $\bm{d}\equiv\left(d_z,d_x\right)$ near the quadratic band touching point $\Gamma$ 
			at fillings $\nu=3/4$ (c) and $\nu=1/4$ (d)
			resembles the vortex in $XY$ systems with the winding number $n=2$.
		}
		\label{fig:band}
	\end{figure}


	We begin with the tight-binding model that describes the hopping process of spinless fermions in the $p$-band honeycomb lattice depicted in Fig.~\ref{fig:band}(a). Introducing an orbital-sublattice spinor representation $p_{\bf k}=\left[p_{x\text{A}{\bf k}},p_{y\text{A}{\bf k}},p_{x\text{B}{\bf k}},p_{y\text{B}{\bf k}}\right]^\text{T}$, the Hamiltonian in the momentum space reads
	\begin{equation}
	\mathscr{H}_{\text{TB}} = \sum_{\bf k} p^\dagger_{\bf k} \mathcal{H}_{\bf k} p_{\bf k},
	\mathcal{H}_{\bf k}=\left[
	\begin{matrix}
	0 & T_{\bf k} \\
	T^\dagger_{\bf k}  & 0
	\end{matrix}
	\right],
	\label{eq:TB}
	\end{equation}
	where
	\begin{equation}
	T_{\bf k}=
	\left[
	\begin{matrix}
	t_\pi+\frac{3t_\sigma+t_\pi}{4}\left(e^{ik_1}+e^{ik_2}\right) & 
	\frac{\sqrt{3}\left(t_\sigma-t_\pi\right)}{4}\left(e^{ik_1}-e^{ik_2}\right) \\
	\frac{\sqrt{3}\left(t_\sigma-t_\pi\right)}{4}\left(e^{ik_1}-e^{ik_2}\right) &
	t_\sigma+\frac{t_\sigma+3t_\pi}{4}\left(e^{ik_1}+e^{ik_2}\right)
	\end{matrix}
	\right]. \nonumber
	\end{equation}
	Here the momenta $k_1$ ($k_2$) is measured along the reciprocal lattice vectors $\bm{b}_1$ ($\bm{b}_2$),
	and the hopping integral $t_\sigma$ ($t_\pi$) denotes the $\sigma$ ($\pi$) bonding of $p$ orbitals. For the $\pi$ bonding, the bond vector lies in the nodal plane of $p$ orbitals. As a result, the strength of $\pi$ bonding is typically much weaker than that of $\sigma$ bonding. The band structure of the tight-binding model in Eq.~(\ref{eq:TB}) with $\{t_\sigma,t_\pi\}=\{1,-0.1\}$, plotted in Fig.~\ref{fig:band}(b), is symmetric with respect to zero energy, arising from the particle-hole symmetry $\mathcal{P}$. Under this symmetry, the tight-binding Hamiltonian is transformed as $\Xi\mathcal{H}_{\bf k}\Xi^{-1}=-\mathcal{H}_{-{\bf k}}$ with the unitary operator $\Xi=s_z\mathcal{K}$. Here $s_z$ is the $z$-component Pauli matrix operating on the sublattice degree of freedom and $\mathcal{K}$ is the complex conjugate operator. 
	The middle two bands cross at the Dirac points located at $K$ and $K^\prime$ points of the hexagonal Brillouin zone (HBZ). While, the lower and upper two bands touch at the $\Gamma$ point of HBZ, and pin the Fermi level at filling $\nu=1/4$ and the particle-hole conjugated filling $\nu=3/4$, respectively. To describe the corresponding low-energy behavior around $\Gamma$ point at filling $\nu=1/4$, we introduce a pseudospin $\sigma_z=\pm1$ to lable the eigenstates $\psi_\Gamma^+=\frac{1}{\sqrt{2}}\left[1,0,-1,0\right]^\text{T}$ and $\psi_\Gamma^-=\frac{1}{\sqrt{2}}\left[0,1,0,-1\right]^\text{T}$. The effective two-band $k\cdot p$ model (see Supplemental Material~\cite{SM} for details) in this basis is given by 
	\begin{eqnarray}
	\mathcal{H}_{\Gamma} \left({\bf k}\right) = 
	d_0\sigma_0+d_x\sigma_x+d_z\sigma_z+\mathcal{O}\left(k^4\right)
	\label{eq:kp}
	\end{eqnarray}
	where $\sigma_0$ is the identity matrix, $\sigma_{x,z}$ are Pauli matrices and the coefficients
	\begin{eqnarray}
	d_0\equiv -\left(t_\sigma+t_\pi\right)&&
	\left[\frac{3}{2}-\frac{3}{8}k^2+\frac{3}{16}\left(\frac{t_\sigma-t_\pi}{t_\sigma+t_\pi}\right)^2k^2\right], \nonumber\\
	\{d_x,d_z\} \equiv \frac{3}{16}&&\left(t_\sigma-t_\pi\right)\{2k_xk_y,k_x^2-k_y^2\}.
	\end{eqnarray}
	Diagonalizing $\mathcal{H}_\Gamma\left({\bf k}\right)$ gives two noninteracting bands $E^\pm_\Gamma\left({\bf k}\right)=d_0\pm\sqrt{d_x^2+d_z^2}$, resulting in a quadratic band touching point (QBTP) at $\Gamma$ point. The pseudovector field $\bm{d}\equiv\left(d_z,d_x\right)$ shown in Fig.~\ref{fig:band}(d) has $d$-wave symmetry. The topological charge for the QBTP is given by the winding number of pseudovector field: $n=\frac{1}{2\pi}\oint_\mathcal{C}\nabla\theta\left({\bf k}\right)\cdot d{\bf k}=2$, where $\theta\equiv\text{arctan}\left(d_x/d_z\right)$ and $\mathcal{C}$ is a contour enclosing the singular $\Gamma$ point, indicating that the QBTP carries a $2\pi$ Berry flux~\cite{Volovik09}.  
	The low-energy Hamiltonian at filling $\nu=3/4$ can be easily derived by applying the particle-hole transformation on the Hamiltonian in Eq.~(\ref{eq:kp}) at filling $\nu=1/4$. The corresponding pseudovector field is shown in Fig.~\ref{fig:band}(c). Hereafter we will only focus on the $\nu=1/4$ filling to simplify the discussion. 	
	
	\begin{figure}
		\centering
		\includegraphics[width=0.48\textwidth]{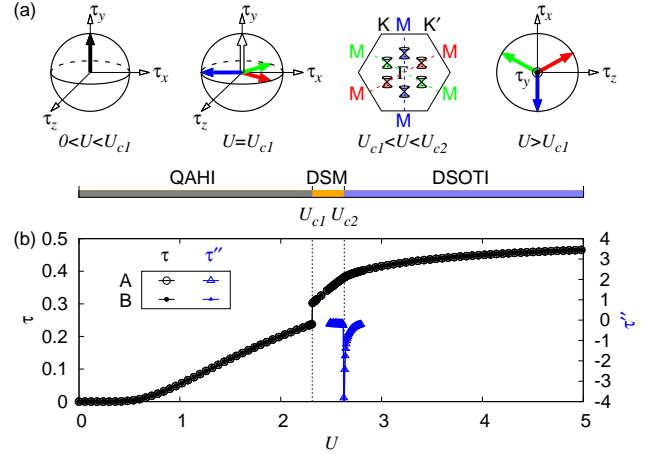}
		\caption{(color online)
			(a) The phase diagram as a function of Hubbard interaction $U$ at $\{t_\sigma,t_\pi\}=\{1,0\}$ shows three phases:
			(1) Quantum anomalous Hall insulator (QAHI) with the spontaneous $y$-axis pseudospin ordering.
			(2) Dirac semimetal (DSM) with the pseudospin ordering aligned with the bond vector $\bm{e}_{1,2,3}$ in $zx$ plane.
			(3) Dimerized second-order topological insulator (DSOTI).
			(b) The evolution of pseudospin magnitude $\tau$. 
			The black dashed line at $U_{c1}$ marks the discontinuous pseudospin-flop transition separates QAHIs from DSMs.
			The black dashed line at $U_{c2}$, indicated by the singular behavior in the second derivative $\tau^{\prime\prime}=d^2\tau/dU^2$, marks the continuous transition between DSMs and DSOTIs.
		}
		\label{fig:hf}
	\end{figure}

	The QBTP is predicted to be generally unstable against many-body interactions towards a broken-symmetry phase by the renormalization group analysis~\cite{Sun09,Zhang12,Herbut14,Murray14}. We are therefore in the position to start from the QBTP and study the phase diagram of present model with Hubbard interactions. Introducing a pseudospin $\tau_z=\pm1$ to label the $p_x$ and $p_y$ orbitals, the Hubbard interaction, at mean-field level, is described by Hartree and pseudospin exchange self-energies
	\begin{equation}
	\mathscr{H}_\text{I}
	=\frac{U}{2}\sum_i\left(n_i\hat{n}_i-\bm{\tau}\hat{\bm{\tau}}-\frac{n_i^2-\bm{\tau}^2}{2}\right)
	\label{eq:HF}
	\end{equation}
	where $\hat{n}_i=\sum_{\mu=x,y}p^\dagger_{i\mu}p_{i\mu}$ and $\hat{\bm{\tau}}=\sum_{\mu\nu}p^\dagger_{i\mu}\bm{\sigma}_{\mu\nu}p_{i\nu}$ are the density and pseudospin operators at the $i$-th site, with $n_i$ and $\bm{\tau}_i$ as the ground-state expectation values. The first term in Eq.~(\ref{eq:HF}) renormalizes the on-site energy level. While, the pseudospin exchange interaction, the second term in Eq.~(\ref{eq:HF}), favors the pseudospin order by lowering the exchange self-energy. As sketched in Fig.~\ref{fig:hf}(a), the calculated phase diagram with the hopping integrals $\{t_\sigma,t_\pi\}=\{1,0\}$ accommodates three different phases including QAHIs, DSMs, and DSOTIs. The effects of $t_\pi$ will be discussed later. Figure~\ref{fig:hf}(b) plots the evolution of pseudospin magnitude $\tau$ for both A and B sublattices. Across the whole phase diagram, the sublattices A and B develop the identical pseudospin order, preserving the inversion symmetry $\mathcal{I}$ with the corresponding unitary operator $I=s_x$. Initially, a weak Hubbard interaction $U$ drives a $y$-axis pseudospin ordering, as a result of spontaneous time-reversal symmetry $\mathcal{T}$ breaking. Consequently, the lower two quadratically touched bands at the HBZ center ($\Gamma$ point) are inverted, accompanied by the degeneracy lifting and the opening of a topological band gap. This insulating phase has non-trivial band topology characterized by the Chern number
	\begin{eqnarray}
	\text{Ch}&=&\int_\text{BZ}d^2{\bf k}\sum_{m\ne n}\left[n_\text{FD}\left(E^n_{\bf k}\right)-n_\text{FD}\left(E^m_{\bf k}\right)\right] \nonumber\\
	&&\times \frac{1}{2\pi} 
	\frac{\text{Im}\left[	\langle n{\bf k}|\hat{v}_x| m{\bf k}\rangle \langle m{\bf k}|\hat{v}_y| n{\bf k}\rangle \right]}
	{\left(E^n_{\bf k}-E^m_{\bf k}\right)^2}
	\label{eq:chern}
	\end{eqnarray} 
	with $n_\text{FD}\left(E\right)$ being the Fermi-Dirac distribution function and $\hat{v}_{\mu}=\partial\mathscr{H}_\text{TB}/\partial k_\mu$ being the velocity operator. In the insulating case, the Hall conductance is determined by the Chern number of the occupied bands and must be an exact integer in the unit of the conductance quantum $e^2/h$~\cite{Klitzing80,Thouless82,Kohmoto85}. Explicit evaluations of Eq.~(\ref{eq:chern}) give $\text{Ch}=\pm1$ as the result of spontaneous time-reversal symmetry $\mathcal{T}$ breaking by freely selecting the pseudospin ordering vector aligned $\mp y$ axis, resembling the $\mathbb{Z}_2$ Ising transition. The nontrivial topological property arises from the orbital angular momentum of the ground state $p_\pm=p_x \pm ip_y$ of a $p$-orbital doublet with its degeneracy lifted by the pseudospin exchange along $\mp y$ axis. Correspondingly, the system has a quantized Hall conductivity $\sigma_{xy}=\pm e^2/h$ in the absence of external magnetic fields and supports a single gapless chiral edge modes (see below). This system is thus identified as a QAHI. We stress that the mechanism of QAHI relies on the spontaneous time-reversal symmetry $\mathcal{T}$ breaking, and is fundamentally different from that of single-particle QAHI~\cite{Wu08}. At the critical Hubbard interaction $U_{c1}\approx2.3$, a pseudospin-flop transition from $y$ axis to $zx$ plane occurs. In the latter phase, the pseudospin vector ${\bm \tau}=\left(\tau_z,\tau_x\right)$ is align with one of the bond vector $\bm{e}_{1,2,3}$, and thus breaks the $C_3$ point group symmetry of honeycomb lattice. As depicted in Fig.~\ref{fig:hf}(a), we find that the band structures in the latter phase hosts a pair of emergent Dirac points each with Berry flux $\pi$, originating from the QBTP with Berry flux $2\pi$ in the noninteracting bands. The transition from QAHI to DSM is expected to be of first order type due to the distinct broken symmetries of these two phases. The corresponding phase boundary is indicated by the discontinuous jump of the pseudospin magnitude $\tau$, reflecting the abrupt pseudospin-flop transition. Upon increasing the interaction $U$, the pair of Dirac points move in the opposite directions towards the $M$ point along the high symmetry line $\Gamma$-$M$ in the HBZ. In Fig.~\ref{fig:hf}(b), a close inspection on the mean-field order parameter reveals that the pseudospin magnitude $\tau$ has a kink around the critical interaction $U_{c2}\approx2.6$, which is best visualized by the second derivative on $\tau$ with respect to the Hubbard interaction $U$, $\tau^{\prime\prime}=d^2\tau/dU^2$. Across the critical interaction $U_{c2}$, the pair of Dirac points approach with each other and annihilate simultaneously at the $M$ point of HBZ, resulting in an insulating phase. The peak of $\tau^{\prime\prime}$, showing a singular behavior, well detects this semimetal-insulator transition. For this insulating phase, an explicit evaluation of the Chern number in Eq.~(\ref{eq:chern}) gives $\text{Ch}=0$, implying that this phase is first-order topological trivial. Below, we will show that this insulator instead is a second-order topological insulator. The polarization for the lowest occupied band ($n=1$) along the primitive lattice vector ${\bm a}_\mu$ is given by
	\begin{equation}
	p_\mu=\frac{1}{\Omega_\text{BZ}}\text{Im}\left[\int_\text{BZ}d^2{\bf k}
	\langle n=1{\bf k}|\hat{\bm b}_\mu\cdot\nabla_{\bf k}|n=1 {\bf k}\rangle \right]
	\label{eq:polar}
	\end{equation}
	with $\Omega_\text{BZ}$ being the area of HBZ and $\hat{\bm b}_\mu$ being the unit reciprocal lattice vector~\cite{Resta07}. The polarization inherits the translation symmetry of Bloch wavefunctions and is uniquely redefined by $\left(p_\mu \text{ mod } 1 \right) \to p_\mu$, representing the shift of Wannier orbital center in units of $\bm{a}_\mu$ away from lattice sites. Moreover, it is straightforward to show that $p_\mu$ is odd under the afore-mentioned inversion symmetry $\mathcal{I}$. Combined with these two symmetries, the polarization $p_\mu$ has a quantized value $0$ or $1/2$ for a gapped system~\cite{Benalcazar17-1,Benalcazar17-2}. In Eq.~(\ref{eq:TB}), we have chosen a gauge such that the hopping processes along the bond vector $\bm{e}_3$ have no phase factor in the unit cell. Therefore, the shift of Wannier orbital center along bond vector $\bm{e}_3$ cannot be captured under this gauge. While, numerical evaluations of $\left(p_1,p_2\right)$ yield $\left(1/2,0\right)$ and $\left(0,1/2\right)$ for the pseudospin vector ${\bm \tau}=\left(\tau_z,\tau_x\right)$ aligned with bond vector $\bm{e}_1$ and $\bm{e}_2$, respectively. In this insulating phase, the Wannier orbital center locates exactly at the center of the corresponding bond to minimize the kinetic energy, supporting zero-dimensional corner states (see below). We thus denote this insulator as a DSOTI with the polarizations as its bulk topological numbers. The DSOTI is a band insulator in nature with one particle occupying a dimerized bond, validating our weak coupling approach here.

	\begin{figure}
		\centering
		\includegraphics[width=0.48\textwidth]{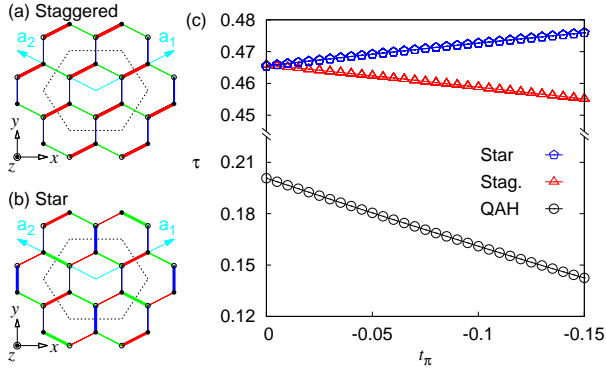}
		\caption{(color online) 
			The pictorial representation of the staggered (a) and star (b) dimerized patterns allowed in the enlarged six-site unit cell. (c) The magnitude of self-consistent pseudospin $\bm{\tau}$ vector with $t_\sigma=1$ as a function of $t_\pi$ for the quantum anomalous Hall phase at $U=2$ and for the staggered and star dimerized patterns at $U=5$.}
		\label{fig:dimer}
	\end{figure}

	Having settled the nature of DSOTIs, we then turn to discuss the effects of the $\pi$ bonding $t_\pi$ on the dimerized bond patterns. Yet, the dimerization was intensively studied in frustrated spin systems~\cite{Rokhsar88,Moessner11,Misguich13}. It is instructive to follow Ref.~\cite{Moessner01} by performing an elemental hexagonal plaquette calculation, whose enlarged unit cell is shown by the dashed line in Fig.~\ref{fig:dimer}(a) and \ref{fig:dimer}(b). We find two stable mean-field solutions, {\it i.e.} the staggered and star patterns, depicted in Fig.~\ref{fig:dimer}(a) and (b), respectively. These two configurations, at mean-field level, are degenerate in energy at $t_\pi=0$. As shown in Fig.~\ref{fig:dimer}(c), the $\pi$ bonding $t_\pi$ reduces (enhances) the pseudospin magnitude $\tau$ of the staggered (star) configuration, thus favoring the star configuration as its ground state in the DSOTI phase. While for the QAHI phase, the $\pi$ bonding $t_\pi$ reduces the corresponding order parameter $\tau$ and thus shifts the phase boundary to a small Hubbard interaction $U$. Numerically, we have verified that both the QAHI phase at $U=2$ and the DSOTI phase at $U=5$ are stable against the DSM phase up to the perturbation $t_\pi=-0.15$. Considering the gapless feature of Dirac semimetal, it deserves further studies with advanced numerical methods, {\it e.g.} quantum Monte Carlo simulations, to examine the effect of quantum fluctuations beyond the mean-field approximation in the future. 

	\begin{figure}
	\centering
	\includegraphics[width=0.48\textwidth]{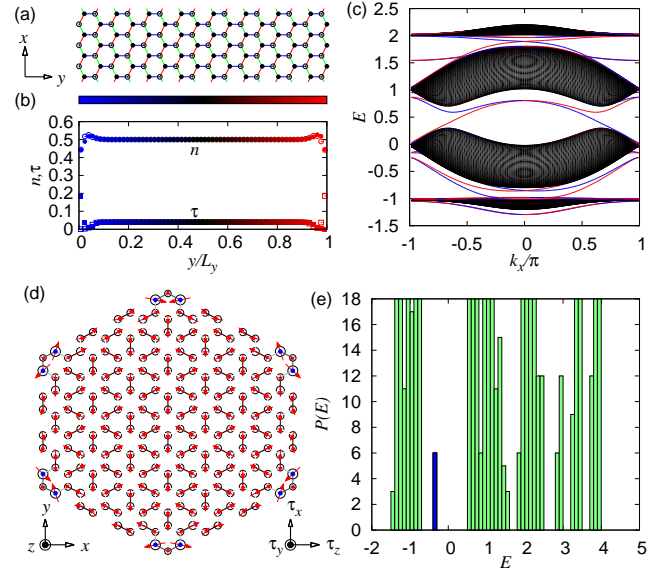}
	\caption{(color online) 
		(a) Schematic plot for a cylindrical geometry with periodic boundary in $x$ direction and open boundary in $y$ direction.
		(b) The self-consistent order parameters for sublattice A (solid symbols) and sublattice B (open symbols) obtained on a cylinder of length $L_y=64$ with the parameters $\{t_\sigma,t_\pi,U\}=\{1,0,2\}$. (c) Energy spectrum of the cylinder in (b). The color encodes the position $\langle y \rangle/L_y$ of the Bloch states illustrated in the color bar in (b). (d) The self-consistent order parameters for the finite-size cluster in the shape of a regular hexagon with the parameters $\{t_\sigma,t_\pi,U\}=\{1,0,5\}$. The black open circle at each site measures the density $n_i$ of the occupied states. 
		The red arrows represent the site-resolved pseudospin vector ${\bm \tau}=\left(\tau_z, \tau_x\right)$. The enhanced bonding energy are highlighted by black solid lines. (e) The histogram measures the distribution of energy levels for the finite-size cluster depicted in (d). The six corner states are labeled in the blue bar. The corresponding spatial distribution $n_i$ of the corner states are represented by the size of the blue solid points in (d).  
	}
	\label{fig:top}
	\end{figure}

	The topgological nature of QAHI and DSOTI is further confirmed by the bulk-boundary correspondence. As shown in Fig.~\ref{fig:top}(a), a cylindrical geometry with two one-dimensional edges along $y$ direction is employed to demonstrate the first-order topological nature of QAHI.	Figures~\ref{fig:top}(b) and \ref{fig:top}(c) show the self-consistently determined order parameters and the energy spectrum, respectively. Both the order parameters $n$ and $\tau$ close to the sample edge deviate from their bulk values, indicating the existence of the edge modes. In the spectrum, each of four bulk bands is split into a series of sub-bands due to the open boundary condition in $y$ direction. Between the lower two complexes of sub-bands, two one-dimensional chiral modes in blue and red colors localize at the right and left edges, respectively. To demonstrate the second-order topological nature of DSOTI, we perform a self-consistent calculation on a finite-size cluster of hexagonal shape with zero-dimensional boundaries, {\it i.e.} six vertices. Figures~\ref{fig:top}(d) and \ref{fig:top}(e) plot the order parameters and the energy spectrum, respectively. Six in-gap bound states, plotted in blue color, are the zero-dimensional corner states arising from the topology of DSOTI. Interestingly, the corner states, localized around six vertices, reside the density to the nearest neighbor sites, alleviating the energy cost due to the Pauli blocking. It is also noteworthy that the $p$ orbitals at the sample edges are spontaneously dimerized in pairs by adjusting the direction of pseudospin vector ${\bm \tau}=\left(\tau_z,\tau_x\right)$, which is dramatically different from the non-interacting case~\cite{Ezawa18}.
	
	In conclusion, the spinless fermions in the $p$-band honeycomb lattice undergo a sequence of Quadratic band touched semimetal-QAHI-DSM-DSOTI transitions with increasing the Hubbard interaction $U$. It is thus remarkable that tuning the Hubbard interaction can switch the system between the first-order and second-order topological insulators beyond the Landau paradigm. Motivated by the experimental advances on $p$-orbital optical lattices~\cite{Bloch07,Wirth11,Soltan-Panahi12,Olschlager13}, we will discuss the possible experimental relevance of our study.
	Experimentally, the honeycomb lattice can be realized by three interfering traveling laser beams~\cite{Grynberg93}, and the Hubbard interaction can be achieved via the $p$-wave Feshbach resonance~\cite{Chin10} according to the Fermi statics~\cite{Chen18}. In particular, ultracold atoms provide various flexible probing techniques~\cite{Bloch08}. The QBTP and Dirac Point can be distinguished by recording the trajectory of transfered atoms across the band degenerate point through the Bloch-Landau-Zener oscillation~\cite{Tarruell12}. In addition, an atomic interferometer can be utilized to measure the Berry flux of Dirac point and QBTP~\cite{Duca15}. The nontrivial Chern number of QAHI can be directly measured by the center-of-mass drift as a Hall response to an external force~\cite{Dauphin13,Aidelsburger14}. Moreover, the DSOTI is easily identified by the single-lattice-site-resolved quantum-gas microscopes~\cite{Bakr09,Sherson10,Bakr10,Gemelke09,Simon11,Omran15,Greif16,Cheuk16,Parsons16,Boll16}. We therefore propose that the topological switch found here can be realized and detected using ultracold atoms in optical lattices.
	
	\paragraph{\it{Acknowledgement.---}}
	We thank X.-J. Liu for helpful discussions. 
	This work is supported by the National Natural Science Foundation of China under Grants No. 11704338, No. 11534001, No. 11504008, and the National Basic Research Program of China under Grant No. 2015CB921102.

	\clearpage
	
	\widetext
	
	\begin{center}
		\textbf{\large Supplemental Material for "Interaction-Driven Topological Switch in a $P$-Band Honeycomb Lattice"}
	\end{center}
	
	
	\appendix
	\section{Derivation of the effective two-band $k\cdot p$ model at filling $\nu=1/4$}
	\label{app:HG}
	
	Introducing the four-component orbital-sublattice spinor representation $p_{\bf k}=\left[p_{x\text{A}{\bf k}},p_{y\text{A}{\bf k}},p_{x\text{B}{\bf k}},p_{y\text{B}{\bf k}}\right]^\text{T}$, 
	the tight-binding model that describes the hopping process of spinless fermions in the $p$-band honeycomb lattice has the following form
	\begin{equation}
	\mathscr{H}_{\text{TB}} = \sum_{\bf k} p^\dagger_{\bf k} \mathcal{H}_{\bf k} p_{\bf k},
	\mathcal{H}_{\bf k}=\left[
	\begin{matrix}
	0 & T_{\bf k} \\
	T^\dagger_{\bf k} & 0
	\end{matrix}
	\right], \nonumber
	\end{equation}
	where
	\begin{equation}
	T_{\bf k}=
	\left[
	\begin{matrix}
	t_\pi+\frac{3t_\sigma+t_\pi}{2}\cos\left[\frac{\sqrt{3}}{2}k_x\right]\exp\left[i\frac{3}{2}k_y\right] & 
	\frac{\sqrt{3}\left(t_\sigma-t_\pi\right)}{2}i\sin\left[\frac{\sqrt{3}}{2}k_x\right]\exp\left[i\frac{3}{2}k_y\right] \\
	\frac{\sqrt{3}\left(t_\sigma-t_\pi\right)}{2}i\sin\left[\frac{\sqrt{3}}{2}k_x\right]\exp\left[i\frac{3}{2}k_y\right] &
	t_\sigma+\frac{t_\sigma+3t_\pi}{2}\cos\left[\frac{\sqrt{3}}{2}k_x\right]\exp\left[i\frac{3}{2}k_y\right]
	\end{matrix}
	\right]. \nonumber
	\end{equation}
	where the hopping integrals $t_\sigma$ and $t_\pi$ denote the $\sigma$ and $\pi$ bonding of $p$ orbitals, respectively.	At the hexagonal Brillouin zone center ($\Gamma$ point), the band dispersions have two sets of two-fold band degeneracy with eigen energies $E_\Gamma^\pm=\pm\frac{3}{2}\left(t_\sigma+t_\pi\right)$, 
	which are exactly the Fermi levels at filling $\nu=3/4$ and $\nu=1/4$, respectively.
	The eigen vectors of the upper eigen energy $E_\Gamma^+$ are given by
	\begin{eqnarray}
	\psi_\Gamma^+\left(+\right) &=& \frac{1}{\sqrt{2}}\left[1,0,1,0\right]^\text{T},\nonumber\\
	\psi_\Gamma^-\left(+\right) &=& \frac{1}{\sqrt{2}}\left[0,1,0,1\right]^\text{T}.\nonumber
	\end{eqnarray}	
	Similarly, the eigen vectors of the lower eigen energy $E_\Gamma^-$ are given by
	\begin{eqnarray}
	\psi_\Gamma^+\left(-\right) &=& \frac{1}{\sqrt{2}}\left[1,0,-1,0\right]^\text{T},\nonumber\\
	\psi_\Gamma^-\left(-\right) &=& \frac{1}{\sqrt{2}}\left[0,1,0,-1\right]^\text{T}.\nonumber
	\end{eqnarray}	
	The upper ($E_\Gamma^+$) and lower ($E_\Gamma^-$) bands at $\Gamma$ point are well separated in energy by a gap $\Delta=3\left(t_\sigma+t_\pi\right)$.
	The low-energy behavior of spinless fermions around $\Gamma$ at filling $\nu=1/4$ is renormalized by a second-order virtual
	process in which the fermion first hops from the lower bands to the upper bands and then hops back to the lower bands. 
	By integrating out these high-energy bands near $\Gamma$ point, 
	the effective two-band $k\cdot p$ model at filling $\nu=1/4$ is given by 
	\begin{equation}
	\mathcal{H}_\Gamma\left({\bf k}\right) = \mathcal{H}_{\bf k}^{--}
	-\mathcal{H}_{\bf k}^{-+}\frac{1}{\mathcal{H}_{\bf k}^{++}-E_\text{F}}\mathcal{H}_{\bf k}^{+-}
	\nonumber
	\end{equation}
	where 
	\begin{equation}
	\mathcal{H}_{\bf k}^{\alpha\beta}=
	\left[
	\begin{matrix}
	\langle \psi_\Gamma^+\left(\alpha\right) | \mathcal{H}_{\bf k} |  \psi_\Gamma^+\left(\beta\right) \rangle & 
	\langle \psi_\Gamma^+\left(\alpha\right) | \mathcal{H}_{\bf k} |  \psi_\Gamma^-\left(\beta\right) \rangle \\
	\langle \psi_\Gamma^-\left(\alpha\right) | \mathcal{H}_{\bf k} |  \psi_\Gamma^+\left(\beta\right) \rangle &
	\langle \psi_\Gamma^-\left(\alpha\right) | \mathcal{H}_{\bf k} |  \psi_\Gamma^-\left(\beta\right) \rangle 
	\end{matrix}
	\right]. \nonumber
	\end{equation}
	and the Fermi level $E_\text{F}=-\frac{3}{2}\left(t_\sigma+t_\pi\right)$.
	After a lengthy but straightforward algebra, the effective Hamiltonian expanded up to quadratic order in ${\bf k}^2$
	takes the form
	\begin{eqnarray}
	\mathcal{H}_{\Gamma} \left({\bf k}\right) = 
	d_0\sigma_0+d_x\sigma_x+d_z\sigma_z+\mathcal{O}\left(k^4\right) \nonumber
	\end{eqnarray}
	where $\sigma_0$ is the identity matrix, $\sigma_{x,z}$ are Pauli matrices and the coefficients
	\begin{eqnarray}
	d_0\equiv -\left(t_\sigma+t_\pi\right)&&
	\left[\frac{3}{2}-\frac{3}{8}k^2+\frac{3}{16}\left(\frac{t_\sigma-t_\pi}{t_\sigma+t_\pi}\right)^2k^2\right], \nonumber\\
	\{d_x,d_z\} \equiv \frac{3}{16}&&\left(t_\sigma-t_\pi\right)\{2k_xk_y,k_x^2-k_y^2\}. \nonumber
	\end{eqnarray}

\end{document}